\documentstyle[prl,aps,twocolumn,epsfig,floats,times]{revtex}

\newcommand  {\rsq}    {\left< r^{2} \right>}

\begin{document}
\draft \title{Self-similar chain conformations in polymer gels}

\author{Mathias P\"utz ${}^{+}$,  Kurt Kremer and
  Ralf Everaers}

\address{Max-Planck-Institut f\"ur Polymerforschung,
  Postfach 3148, D-55021 Mainz, Germany}

\address{$+$ present address:
  Sandia National Laboratories,
  Albuquerque,NM 87185-1349, USA}
\address{
\begin{minipage}{5.55in}
\begin{abstract}\hskip 0.15in
We use molecular dynamics simulations to study
the swelling of randomly end-cross-linked polymer networks
in good solvent conditions. We find that the equilibrium degree
of swelling saturates at $Q_{eq}\approx N_e^{3/5}$ for
mean strand lengths $\bar{N_s}$ exceeding the melt entanglement length $N_e$.
The internal structure of the network strands in the swollen state
is characterized by a new exponent $\nu=0.72\pm0.02$.
Our findings are in contradiction to de Gennes' $c^\ast$-theorem,
which predicts $Q_{eq}\sim N_s^{4/5}$ and $\nu=0.588$.
We present a simple Flory argument for 
a self-similar structure of mutually interpenetrating
network strands, which yields $\nu=7/10$ and otherwise
recovers the classical Flory-Rehner theory. In particular,
$Q_{eq}\approx N_e^{3/5}$, if $N_e$ is used
as effective strand length.
\end{abstract}
\end{minipage}
\vspace*{-0.5cm} 
} 
\maketitle

Polymer gels~\cite{Flory_53,deGennes_79,Derosi_91,Dusek_advpolsc_93,Addad_96}
are soft solids governed by a complex interplay of the elasticity of
the polymer network and the polymer/solvent interaction.
They are sensitive to the preparation conditions 
and can undergo large volume changes in response to small variations of a 
control parameter such as temperature, solvent composition, pH or 
salt concentration. In this letter we reexamine a classical but
still controversial problem of polymer physics~\cite{Flory_53,deGennes_79}, 
the equilibrium swelling of a piece of rubber in good solvent.

Experimentally gels have been studied extensively by
combining thermodynamic and rheological investigations 
with neutron or light scattering~\cite{Tanaka_jcp_73,Candau_advpolsc_82,%
Bastide_Addad_96}.
Here we use computer simulations~\cite{Kremer_Binder94,Trautenberg_faraday_95,%
REKK_jmolmod_96,HTG_prl_97,Escobedo_jcp_96}, since they offer some advantages 
in the access to and the control over microscopic details of the network 
structure. We concentrate on the role of 
entanglements in limiting the swelling process and, in particular, 
the structure of the network strands in the swollen gel.
Questions relating to the structural heterogeneity in our gels
and the butterfly effect~\cite{Bastide_Addad_96,PanyukovRabin_pr_96}
will be addressed in a future publication.      

There are several theories for the swelling of polymer networks 
prepared from a melt of linear precursor chains. 
In the dry state of preparation, the network strands
have Gaussian statistics, i.e. the mean square end-to-end distance
is related to the average length, $\bar{N_s}$,
by $\rsq_{dry} \propto b^2 \bar{N_s}^{2\nu}$, 
where $\nu=1/2$ and $b$ is the 
monomer radius. The same  relation also holds for all internal
distances, leading to the characteristic structure factor
$S(q) \sim q^{-1/\nu}$ for the scattering at wave vector $q$
from a fractal object. 

The classical Flory-Rehner theory~\cite{Flory_53} writes the
gel free energy $F$ as a sum of two independent terms: 
a free energy of mixing with the solvent (favoring swelling and
estimated from the Flory-Huggins theory of semi-dilute
solutions of linear polymers) and an elastic free energy
(due to the affine stretching of the network strands which are
treated as Gaussian, concentration-{\em independent},
linear entropic springs).  Minimizing $F$ yields 
$Q_{eq} \propto \bar{N_s}^{3/5}$ for the equilibrium degree
of swelling. 
The Flory-Rehner theory implies that the structure factor of long paths
through the network is of the form $S(q) \sim q^{-2}$
both locally, where the chains are unperturbed, and on large scales,
where they deform affinely 
($\rsq_{eq} \propto \rsq_{dry} Q_{eq}^{2/3}$) with the outer
dimensions of the sample. The stretching should be visible
in the crossover region around $q\approx 2\pi/(b \bar{N_s}^{1/2})$ 
with $S(q) \sim q^{-1}$.

More recent treatments are based on the scaling theory of 
semi-dilute solutions of linear polymers~\cite{deGennes_79}.
Locally, inside of so called ``blobs'', the chains behave 
as isolated, {\em self-avoiding} walks with 
$\nu\approx3/5$. A blob  containing $g$ monomers has  
a typical diameter $\xi\propto b g^\nu =  b Q^{3/4}$.  
Chains with $N\gg g$ can again be regarded as ideal,
however, with a renormalized chain length $N/g$ and a
renormalized monomer size $\xi$. 
Quite interestingly, refining the Flory-Rehner ansatz
along these lines \cite{Panyukov_jetp_90,Obukhov_mm_94,PanyukovRabin_pr_96} 
recovers the classical result $Q_{eq} \propto N_s^{3/5}$.
Such models imply for the structure factor a crossover from 
$S(q)\sim q^{-5/3}$ to $S(q)\sim q^{-1}$ at $q\approx 2\pi/\xi_{eq}$,
where the blob diameter at equilibrium swelling is given by 
$\xi_{eq} \propto b Q_{eq}^{3/4}$ and much smaller than
the strand extension.

An open point is the length scale on which the systems begin
to deform affinely with the macroscopic strain. The two theories
mentioned above consider linear entropic springs,
where due to the global connectivity (disregarding fluctuation effects)
this length scale is given by the strand size. 
A drastically different point of view has been advanced by de Gennes,
who argues that the swelling is limited by the {\em local}
connectivity, which only begins to be felt at
the overlap concentration $c^\ast \propto \bar{N_s}/(b \bar{N_s}^\nu)^3$ 
of a semi-dilute solution of linear polymers of average length $\bar{N_s}$,
corresponding to $Q_{eq}\propto \bar{N_s}^{4/5}$.
As a motivation for his $c^{\ast}$--theorem~\cite{deGennes_79}, 
de Gennes considers {\em crosslinking in dilute solution}, 
but postulates that the same results also hold for 
{\em swelling of networks} prepared by cross-linking dense melts.
The $c^{\ast}$--theorem predicts $S(q)\sim q^{-5/3}$
for $q> 2\pi/(b \bar{N_s}^{3/5})$ as well as unusual elastic properties
due to the {\em non-linear} elasticity
of the network strands~\cite{HTG_prl_97,Daoudi_jphysfr_77}.

Both, the Flory-Rehner theory~\cite{Flory_53} and
the $c^\ast$-theorem~\cite{deGennes_79}
are supported by part of the  
experimental evidence~\cite{Candau_advpolsc_82,Patel_mm_92,Hild_pol_97}. 
While the results are very sensitive to the details of the preparation
process, it seems well confirmed~\cite{Hild_pol_97}
that highly cross-linked networks
behave according to Flory's prediction 
$Q_{eq} \propto N_s^{3/5}$.
On the other hand, SANS 
experiments~\cite{Candau_advpolsc_82,BeltzungPicot_mm_83}
and computer simulations~\cite{Trautenberg_faraday_95} of
lightly cross-linked gels show 
the weak dependence of the strand extensions on the
degree of swelling predicted by the $c^\ast$-theorem.
To our knowledge, all SANS studies have concentrated
on the low $q$ Guinier regime (i.e. the radius of gyration of the strands)
and no particular attention has been paid 
to the local chain structure.

As in earlier investigations of polymer melts and
networks \cite{KremerGSG_jcp_90,GrestKremer_mm_90,DueringKKGSG_prl_91,DueringKKGSG_jcp_94,REKK_mm_95}, we used a coarse-grained polymer model
where beads interacting via a truncated,
purely repulsive Lennard-Jones (LJ) potential are
connected by FENE springs. With $\epsilon$,
$\sigma$ and $\tau$ as the LJ units of energy, length and
time, the equations of motion were integrated by a velocity-Verlet algorithm
with a weak local coupling to a heat bath at $k_B T = 1\epsilon$.
The potentials were
parametrized in such a way that chains were effectively uncrossable,
i.e. the network topology was conserved for all times.
In our studies we did not simulate the solvent
explicitly, but rather used vacuum which can be considered as a perfect
solvent for our purely repulsive (athermal) network chains.
The relevant length and time scales for chains in a melt are 
the average bond length, $\sqrt{ \langle l^2 \rangle}=0.965(5) \sigma$,
the mean-square end-to-end
distance $\langle r^2 \rangle(N)_{dry} = 1.74(2) l^2 N$
\cite{KremerGSG_jcp_90}, the melt entanglement length, $N_e
= 33(2)$ monomers, and the Rouse time $\tau_{Rouse}(N) \simeq 1.35\tau
N^2$ \cite{PuetzKKGSG_prl_99}. In dilute solutions,
single chains adopt self-avoiding conformations with
$\langle r^2 \rangle(N) \approx 1.8 l^2 N^{3/5}$.

Using this model, it is possible to study 
different network structures including randomly cross-linked,
randomly end-cross-linked~\cite{GrestKremer_mm_90,DueringKKGSG_prl_91} 
and end-linked melts~\cite{DueringKKGSG_jcp_94} as well as networks
with the regular connectivity of a crystal lattice~\cite{REKK_mm_95}. 
Here we investigate end-cross-linked
model networks created from an equilibrated monodisperse
melt with $M$ precursor chains of length $N$ at a melt-like
density $\rho_{dry}=0.85 \sigma^{-3}$ by connecting the
end monomers of the chains to a randomly chosen adjacent monomer
of a different chain. This method yields defect-free tri-functional
systems with an exponential distribution of strand lengths $N_s$
with an average of $\bar{N_s}=N/3$. The Gaussian statistics of the strands
remains unperturbed after
crosslinking~\cite{GrestKremer_mm_90,PuetzKK_prep_99}.
The systems studied range from $M/N=3200/25$ (i.e. the average
strand size $\bar{N_s}=8.3$) up to $M/N=500/700$ ($\bar{N_s}=233$),
some systems being as large as $M \cdot N = 5 \cdot 10^5$.
All simulations used periodic boundary conditions in a cubic box
and were performed at constant volume. Starting from 
$V_{dry}=M N/\rho_{dry}$, the size of the simulation box was increased
in small steps alternating with equilibration periods of at least 5
entanglement times $\tau_R(N_e)\approx1400\tau$. The isotropic pressure
$P$ was obtained from the microscopic virial tensor and the condition
$P_{eq}\equiv0$ was used to define equilibrium swelling 
with $Q_{eq}=V_{eq}/V_{dry}$. Tests with a part of the networks using
open boundaries did not show any significant changes of the results.

\begin{figure}
  \begin{center}
    \epsfig{file=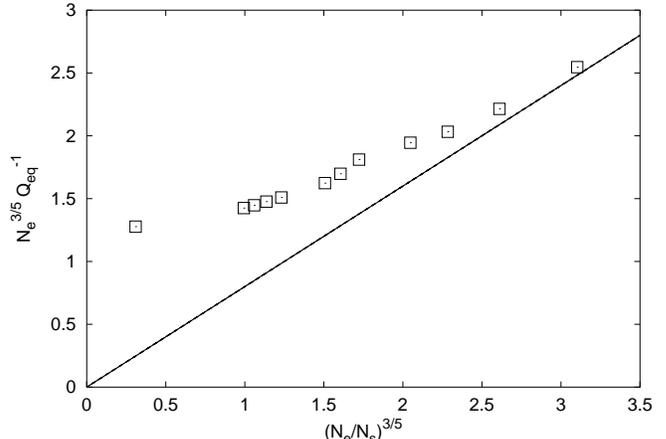,width=8.6cm}
    \caption{ \label{fig:maxswell} 
        Strand length $N_s$  dependence of the equilibrium degree of
        swelling $Q_{eq}$. The straight line going through zero 
        represents Flory's prediction $Q_{eq} \propto \bar{N_s}^{3/5}$.
        The melt entanglement length $N_e$
        was used to normalize the axis in order to show that
        deviations from Flory's theory occur around $N_s\approx N_e$
        and that the asymptotic value $Q_{eq}(\bar{N_s}\rightarrow\infty)$
        is of the order of $N_e^{3/5}$.}
  \end{center}
\end{figure}

We investigated the equilibrium swelling of our model networks
as a function of the average strand length $N_s$.
Fig.~\ref{fig:maxswell} shows $Q_{eq}^{-1} N_e^{3/5}$
as a function of the average strand length $(N_e/N_s)^{-3/5}$.
Our results for short strands are compatible with 
the Flory-Rehner~\cite{Flory_53} prediction $Q_{eq} \propto \bar{N_s}^{3/5}$,
but do not allow for an independent determination of the
exponent. In contradiction to this theory
we observe a saturation of the equilibrium swelling degree for large
$\bar{N_s}$. The crossover occurs for $\bar{N_s} \approx N_e$.
The extrapolated maximal degree of
equilibrium swelling $Q_{max}(\bar{N_s} \rightarrow \infty) = 6.8(3)$
is close to the swelling degree of an ideal Flory-gel with average strand
length $N_e$ : $1.15 \cdot N_e^{3/5}=9.5$, where the prefactor is empirically
obtained from the slope of the straight line in Fig.~\ref{fig:maxswell}.
In contrast, the corresponding estimate based on the
$c^\ast$-theorem, $Q_{eq} \simeq b^3/\sigma^3 N_e^{4/5} \approx 36$,
is clearly too high ($b=1.3\sigma$ is the stastical segment length 
in good solution).
Our interpretation is that to a first approximation
entanglements act as chemical crosslinks in limiting the
swelling of polymer networks. The situation is analoguos to 
an ``olympic gel''\cite{deGennes_79} of topologically linked 
ring polymers. In contrast to solutions of
linear polymers, systems containing trapped entanglements
cannot be arbitrarily diluted.

\begin{figure}
  \begin{center}
 \epsfig{file=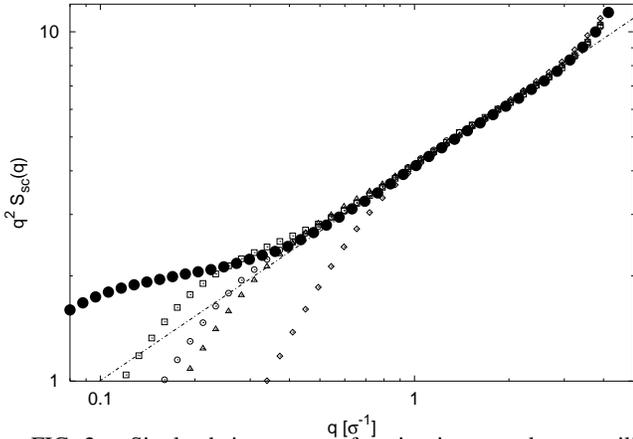,width=8.6cm}
    \caption{ \label{fig:singlestr} Single chain structure function
      in networks at equilibrium swelling in Kratky representation. 
      The straight line corresponds to a power law $q^{2-1/\nu}$ 
      ($\nu = 0.72$). 
      The figure contains scattering data for the precursor chains
      ($\bullet$) of length $N=700$ ($\bar{N_s}=233$) and for network
      strands of lengths $N_s=10$ ($\Diamond$), $N_s=40$ ($\triangle$),
      $N_s=70$ ($\odot$), $N_s=100$ ($\Box$) within a different network
      with precursor chains of length $N=100$ ($\bar{N_s}=33$).
      }
  \end{center}
\end{figure}

The chain conformations at equilibrium swelling
are best characterized by their structure factor $S(q)$.
Fig. \ref{fig:singlestr} shows $S(q)$ of the precursor chains
within the network for our most weakly crosslinked $N=700$ sample.
We have chosen the Kratky-representation ($q^2S(q)$ vs. $q$) to show
the deviation from the Gaussian case ($S(q) \propto q^{-2}$) more clearly.
The observed power law form $S(q) \propto q^{-1/\nu}$ 
is characteristic of fractals and common in polymeric systems. 
However, the observed exponent $\nu=0.72(2)$ is unexpected.
Furthermore, the fractal structure is observed for a 
$q$-range of $2\sigma \lesssim \frac{2\pi}q \lesssim 15.5\sigma
\approx b N_e^{0.72}$, suggesting that the 
mean extension of the effective strands of length $N_e$
is the only relevant length scale in the problem.
For smaller $q$ we see the onset of the expected scattering
of a Gaussian chain consisting of randomly oriented parts of
length $b N_e^{0.72}$. Our precursor chains (even $N=700$)
are to short to see it clearly developed.

Since the scattering from the {\em precursor chains} could be
affected by polydispersity effects,
we have investigated the conformations of the {\em network strands} 
as a function of their contour length $N_s$.
For high $q$ all structure functions fall on top of each other
and show the same fractal structure with $S(q) \sim q^{-1/0.72(2)}$
(Fig.~\ref{fig:singlestr}).
\begin{figure}
  \begin{center}
    \epsfig{file=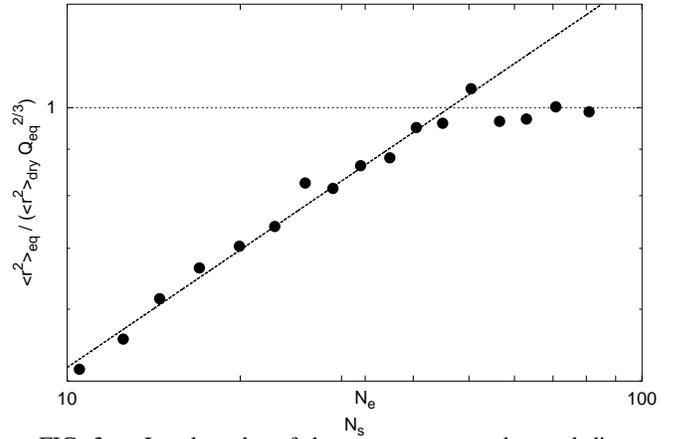,width=8.6cm}
    \caption{ \label{fig:strandsize} Log-log plot of the mean-square
      end-to-end distance $\langle r^2 \rangle_{eq}$ of the individual
      network strands within a single network ($\bar{N_s}=33$) at equilibrium
      swelling $Q=5.8$ versus strand length $N_s$. The straight line
      corresponds to a power law $\rsq_{eq} \sim N_s^{2\nu}$ with $\nu=0.72$. 
      The data are normalized to an affine
      deformation $\rsq_{eq} = Q_{eq}^{2/3} \rsq_{dry}$.}
  \end{center}
\end{figure}
The complementary Fig. \ref{fig:strandsize} shows a log-log plot
of the mean square strand extension $\rsq_{eq}(N_s)$
versus their length. In agreement with the results for
the structure functions, we find a power law 
$\rsq_{eq}\propto  b^2 N_s^{2\times0.72}$ for strands which are shorter
than the effective strand length $N_e$ and therefore {\em sub-affine}
deformations. Long strands, on the other hand, deform affinely
with $\rsq_{eq} = \rsq_{dry} Q_{eq}^{2/3}$.

Clearly, the results of our simulations do not agree with the
predictions of any of the theories presented in the introduction.
While the neglect of entanglements 
seems to be fairly simple to repair by treating them 
as effective cross-links (with $N_e$ supplanting the average strand
length $\bar{N_s}$ \cite{Obukhov_mm_94}), 
the fractal structure of the strands
and the exponent $\nu=0.72(2)$ come as a surprise.
In the following, we discuss a possible explanation
for the stronger swelling of network strands ($\nu\approx 0.72$) 
than of single chains ($\nu\approx 3/5$) under good solvent
conditions.

We begin by recalling Flory's argument~\cite{Flory_53} 
for the typical size $R_F\propto b N^{\nu}$
of a single polymer chain of length $N$ and statistical segment size $b$
in a good solvent. The equilibrium between an elastic energy 
$\propto R_F^2/(b^2 N)$ of a Gaussian chain stretched to $R_F$
and a repulsive energy $\propto b^d R_F^d (N/R_F^d)^2$
due to binary contacts between monomers in $d$ dimensions
leads to

\begin{equation}
\nu=\frac3{d+2}\ \ \ .
\label{eq:RF}
\end{equation}
The simplest models for swollen networks have the regular
connectivity of a crystal lattice. In agreement with the
$c^{\ast}$-theorem they adopt equilibrium conformations 
with strand extensions of the order of $R_F$~\cite{Trautenberg_thesis_97}. 
However, these systems are hardly good models for the swelling process
of networks prepared in the dry state, since the
hypothetical initial state at melt density has an
unphysical local structure with average strand extensions
$R_F Q^{-1/3} \propto b N_s^{1/3}$ as in dense globules. 
In contrast, if the corresponding semi-dilute solution is
compressed, the chains shrink only weakly from $R_F$ to 
the Gaussian coil radius $R \propto b N_s^{1/2}$. Instead, they 
become highly interpenetrating with $n\propto N_s^{1/2}$
of them sharing a volume of $R^3$. Moreover, at least
the simplest model for highly cross-linked
networks prepared in the dry state,
$n\sim N_s^{1/2}$  {\em mutually interpenetrating} 
regular networks with strand extensions of the order of $R$
\cite{REKK_mm_95}, 
cannot possibly comply with the $c^{\ast}$-theorem, if one
disregards {\em macroscopic} chain separation:
Either the strands extend to $R_F$,
leading to internal concentrations of $c^{\ast} N_s^{1/2}$, 
or the systems swell to $c^{\ast}$, in which case 
the strands are stretched to $R_F N_s^{1/6}$. 
The same conclusions should hold for 
any network without too many defects, where the
{\em global} connectivity forces neighboring chains to share the same volume
{\em independent} of the degree of swelling. 

The consequences can be estimated using a simple
Flory argument. Instead of a single polymer, 
we now consider a group of chains which 
can swell but {\em not} desinterpenetrate, i.e.
$n\sim N^{1/2}$ chains of length $N$ which span a volume $R_{FR}^3$.
The equilibrium between the elastic energy 
$\propto n\,R_{FR}^2/(b^2 N)$  and the repulsive energy 
$\propto b^d R_{FR}^d (n\,N/R_{FR}^d)^2$
leads to
\begin{eqnarray}
\nu &=& \frac{4+d}{4+2d}
\label{eq:RFR}
\end{eqnarray}
Quite interestingly, this {\em local} argument reproduces
in three dimensions with 
$Q_{eq}\sim N^{d/(d+2)} = N^{3/5}$ 
and $R_{FR} \sim Q^{1/d} b N^{1/2} \sim N^{7/10}$
the results of the classical Flory-Rehner theory of gels. 
However, in analogy to the Flory argument
for single chains, Eq.~(\ref{eq:RFR}) should also apply
to subchains of length $G$ with $1 \ll G < N$ which share their volume with
a correspondingly smaller number of other subchains.
In particular, the local degree of swelling, $G^{1/5}$, 
should be {\em sub-affine} and {\em the exponent $\nu=7/10$ 
should characterize the entire local chain structure}
up to the length scale of the effective strand length, $N_e$.
This is in excellent agreement with the main findings from
our simulations (see Figs. \ref{fig:singlestr} and \ref{fig:strandsize}).

Before we conclude, some additional remarks are in order:
(i) For swelling in a Theta-solvent, the analogous scaling 
argument yields $Q\sim N^{3/8}$ in agreement with
previous theories and experiments
\cite{Patel_mm_92,Hild_pol_97,Obukhov_mm_94} and predicts
local chain structures characterized by $\nu=5/8$.
(ii) Eq.~(\ref{eq:RFR}) can also be derived 
along the lines of \cite{Panyukov_jetp_90,Obukhov_mm_94,PanyukovRabin_pr_96}
from an equilibrium between the elastic energy of blob chains and 
the osmotic pressure of a semi-dilute polymer solutions. Note, that the
apropriate blob size is a function of the size $G$ of the
subchains under consideration and that isolated chain behavior is
only expected below the original correlation length $\xi_{prep}$ for
systems prepared by cross-linking semi-dilute solutions. 
(iii) Sommer, Vilgis and Heinrich~\cite{Sommer_jcp_94} 
have argued that the effective inner fractal dimension
$d_i$ of a polymer network is larger than $d_i=1$ for 
linear chains, leading to stronger swelling with
$\nu=\frac{d_i+2}{d+2}$. 
While the correction goes into the right direction, 
it is difficult to explain a strand length {\em independent}
effective inner fractal dimension of $d_i=1.5$ as an effect
of the {\em local} connectivity.
(iv) However, such effects may well be important in systems
with a sufficient number of defects such as dangling ends or clusters.
If the global connectivity is weak, the chains may locally desinterpenetrate,
leading to a behavior which agrees much better with the 
$c^\ast$-theorem~\cite{Candau_advpolsc_82,Trautenberg_faraday_95}. 

In summary, we have used large scale computer simulations and
scaling arguments to investigate the swelling behavior of 
defect free model networks prepared at melt density.
We find that for networks with average strand lengths $\bar{N_s}>N_e$ the
swelling is limited by entanglements to $Q_{eq}\approx N_e^{3/5}$
and that the strands locally exhibt a fractal structure characterized by an
exponent $\nu\approx7/10$ which should be directly 
observable in neutron scattering experiments. 

We acknowledge the support of the H\"ochst\-leistungs\-rechenzentrum J\"ulich
and the Rechenzentrum of the MPG in M\"unchen and thank G.~S.~Grest for
discussions and a careful reading of the manuscript.

\bibliographystyle{prsty}

\end{document}